\documentclass{aa_old}  
\usepackage{graphicx}
\usepackage{natbib}
\bibpunct{(}{)}{;}{a}{}{,}
\usepackage{txfonts}
\def\sU{\sigma_{\rm U}}
\def\sU'{\sigma_{\rm U'}}
\def\sV{\sigma_{\rm V}}
\def\sV'{\sigma_{\rm V'}}

\begin{document}
\titlerunning{The Hyades stream}
  \title{The Hyades stream: evaporated cluster or \\ intrusion from
  the inner disk?}   

   \author{B. Famaey
          \inst{1}
          \and
          F. Pont
          \inst{2}
          \and
          X. Luri
          \inst{3}
          \and
	  S. Udry
          \inst{2}
          \and
	  M. Mayor
          \inst{2}
          \and
          A. Jorissen
          \inst{1}
          }

  \offprints{B. Famaey}

   \institute{Institut d'Astronomie et d'Astrophysique, Universit\'e
Libre de
  Bruxelles, CP 226, Boulevard du Triomphe, B-1050 Bruxelles, Belgium            
         \and
            Observatoire de Gen\`eve, Chemin des Maillettes 51, CH-1290
	     Sauverny, Switzerland
         \and
            Departament d'Astronomia i Meteorologia, Universitat de
	     Barcelona, Avda. Diagonal 647, E-08028 Barcelona, Spain
  }

   \date{Received ...; accepted ...}

 
  \abstract
   {
The nature of the Hyades stream, or Hyades moving group, is a
  long-standing question of Galactic Astronomy. While it has become widely
  recognized that the Hercules stream, an unbound group of stars lagging behind
  galactic rotation and moving outward in the galactic disk, is associated
  with the outer Lindblad resonance of the rotating galactic bar, there is
  still some debate about the nature of the more prominent low-velocity
  stream sharing the kinematics of the Hyades open cluster. Is this stream
  caused by additional non-axisymmetric perturbations of the galactic
  potential, such as transient or quasi-stationary spiral waves, or by the
  on-going evaporation of the Hyades cluster?
A simple observational test has been designed to answer that question,
  i.e. to determine whether the
  Hyades stream is primarily composed of coeval stars originating from the
  Hyades cluster, or of field stars. 
Using the Geneva-Copenhagen survey of F and G dwarfs, we compare the mass 
distribution and metallicity of the stream to those of field disk stars. 
If the Hyades stream is composed of stars trapped at resonance, its mass
distribution should obey the present-day mass function (PDMF) of the disk,
and its metallicity should reflect its origin in the inner regions of the
Galaxy. On the other hand, if it is an evaporated cluster, we expect a
different mass distribution, depending on the inital mass function (IMF) of
the cluster, and on the proportion of evaporated stars as a function of mass.
We find that extreme conditions have to be adopted for the selective
  evaporation and IMF of the cluster to make the observed
  mass ditribution of the stream only roughly consistent (at a one-sigma
  level) with the coeval evaporated cluster scenario. The observed mass
  distribution is in much better
  agreement with the PDMF of the field. We also note that the peculiar
  metallicity of the stream is inconsistent with that of a field population
  from the solar neighbourhood trapped in the primordial cluster during its
  formation process and subsequently evaporated. These observations thus
  favour a resonant origin for the Hyades stream, as suggested in Famaey et
  al.\ (2005).  
   \keywords{Galaxy: kinematics and dynamics -- evolution -- clusters and
               associations -- disk -- solar neighbourhood -- stars: kinematics
               }
}
   \maketitle
%

\section{Introduction}

It has been 
known for a very long time that a spatially unbound group of stars in
the solar neighbourhood is sharing the same kinematics as the Hyades open
cluster (Hertzsprung 1909, Str\"omberg 1922, Eggen 1958, Perryman et
al. 1998). Assuming that it
was a vestige of an initially more massive cluster which partly evaporated
with time, Eggen christened this kinematically cold group the Hyades
supercluster. More generally, it is called the Hyades stream, or Hyades moving
group.

During the last fifteen years, Eggen's hypothesis that kinematic groups of
this type are 
in fact cluster remnants has been largely debated, because they may also
be generated by a number of global dynamical mechanisms. A rotating bar at
the centre of the Milky Way may e.g. cause
the velocity distribution in the vicinity of the outer Lindblad resonance
to become bimodal, due to the coexistence of orbits elongated along and
perpendicular to the bar's major axis. Today, this mechanism is thought to actually account for the Hercules
stream (Dehnen 2000), a group of stars lagging behind the galactic
rotation and moving outward in the disk. Moreover, the
perturbation that the triaxial bar induces on a flat axisymmetric disk does
produce some chaos: Fux (2001) showed
that when the bar is taken into account, the chaotic
regions, decoupled from the regular regions, are more
heavily crowded in the region of the Hercules stream in velocity
space. On the other hand, other streams, including the Hyades, could still be linked with the spirality of the
Galaxy, because any perturbation of the axisymmetric potential is likely
to buffet the stars.

Eggen's scenario and the dynamical perturbations are, of course, not
mutually incompatible phenomena. Clusters are
known to evaporate over time, and there must be some intermediate state when
a group of stars no longer spatially identifiable still share similar velocities.
Indeed, disk stars (most of which move
on quasi-circular epicyclic orbits) which formed at the same place and
time, and which stayed in the same region of the Galaxy after a few galactic
rotations must necessarily have the same period
of revolution around the Galactic center, and thus the same guiding-center
which in turn implies 
the same tangential velocity (Woolley 1961). On the other hand, our Galaxy is
known to have large spiral perturbations that are likely to have a
kinematic effect on the velocity of stars. In the past, the importance of stirring by spiral structure has been
underestimated because it was thought that spirals heated the disk
strongly, and because the amount of heating is observationally
constrained. However, Sellwood \& Binney (2002) showed that the dominant
effect of spirals is to stirr without heating.  Thus, the Hyades stream may well be an outward-moving
stream of stars on horseshoe orbits that cross the corotation of the spiral
pattern. The kinematics of K and M giant stars in the solar
neighbourhood seem to comfort this scenario, and suggest that the Hyades
stream is composed of metal-rich stars with a wide range of ages (Famaey et
al. 2005, see also Chereul \& Grenon 2001). However, metallicities were
available for too few giants, and individual age estimates were not precise
enough to reach a decisive conclusion. This motivates the present analysis,
where we examine whether properties of F and G dwarfs from the
Geneva-Copenhagen survey (Nordstr\"om et al. 2004) are in better agreement
with this dynamical scenario, with Eggen's scenario, or with a mix of both.   

\section{The sample}
\begin{figure}
   \centering
   \includegraphics[width=7cm]{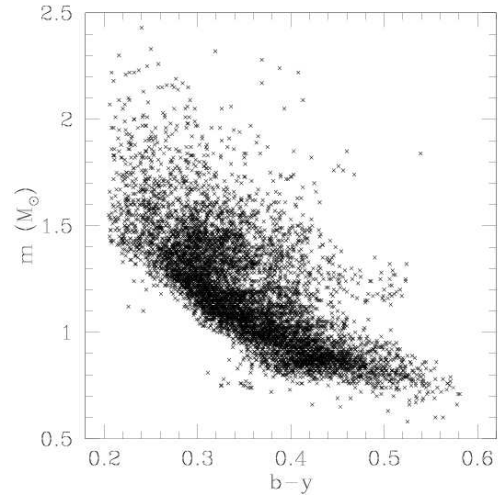}
      \caption{\label{Fig:byM}
The distribution of stars from the Geneva-Copenhagen survey in the ($b-y$ --
mass) plane.
              }
         \label{Fig:members}
   \end{figure}

The Geneva-Copenhagen catalogue of F and G dwarfs\ (Nordstr\"om et
al. 2004) is the result of a decade-long campaign of Str\"omgren photometric
and Coravel spectroscopic measurements for more than 16000 late-type dwarfs
in the solar neighbourhood. It gives positions, parallaxes, proper motions,
masses and metallicities for a sample of dwarfs complete to 40--70 pc
depending on spectral type. 
 Metallicities are determined from Str\"omgren photometry according to
 a tailored calibration described in Nordstr\"om et al. 
(2004; their Sect.~4.3).
Masses are determined by comparing 
the observed position in the Hertzsprung-Russell diagram 
(effective temperatures and absolute magnitudes were derived from
Str\"omgren photometry and Hipparcos parallaxes)
with stellar-evolution tracks from Girardi et al. (2000). Masses are
displayed as a function of the $b-y$ color index in Fig.~\ref{Fig:byM},
which shows that the Geneva-Copenhagen survey is only complete in the
  mass range $[0.8,1.5]$~$M_\odot$.

The survey contains 122 stars flagged as definite or possible members of
  the Hyades cluster by Perryman et al. (1998) and de Bruijne et
  al. (2001). To avoid contamination from the Hyades {\it cluster}
  in our subsequent analysis of the Hyades {\it stream} we exclude those
  122 stars from the sample. After also excluding the binaries to avoid the
  distorting effect of binarity on the photometry, the total
  number of (single) stars in the present sample is 8084.

\section{The velocity field}

  \begin{figure}
   \centering
   \includegraphics[width=7cm]{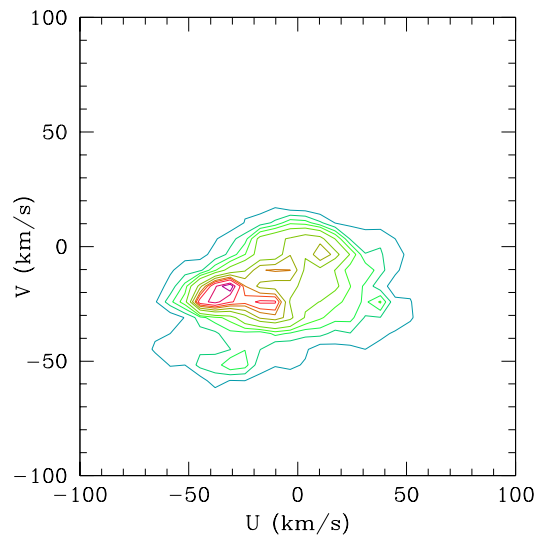}
      \caption{Isocontours for the whole survey in the $UV$-plane: the
      contours correspond respectively to 0.5, 0.8, 1.2, 1.5, 1.9,
              2.6, 3.1, 3.5, 3.8, 4.2, 4.7, 5. stars/(km/s)$^2$. The Hyades
              stream, at $U \simeq -37$~km/s and $V \simeq -17$~km/s, is
              the most prominent feature. 
              }
         \label{Fig_UVfield}
   \end{figure}

 Fig.~\ref{Fig_UVfield} 
presents the distribution of all stars from the
 survey in the $UV$-plane ($U$ is the velocity towards the galactic center,
 $V$ the velocity in the direction of Galactic rotation, both with respect to
 the Sun). For computing $U$ and $V$, distances are adopted from the
 Geneva-Copenhagen Catalogue. These distances are derived from Hipparcos 
parallaxes if their relative errors are better than 13\%, and from photometric 
distances  (which are uncertain by 13\%) otherwise.  

 The small-scale structure in the $UV$-plane
appears clearly on Fig. 2: the main overdensities are the well-known 
Hercules stream
 ($U \simeq -30$~km/s, $V \simeq -50$~km/s), the Sirius-UMa stream ($U
\simeq 10$~km/s, $V \simeq -5$~km/s), the Castor group ($U \simeq -10$~km/s, $V \simeq -10$~km/s), the
 Pleiades stream ($U \simeq -15$~km/s, $V=-25$~km/s), and, the most prominent, the Hyades stream
 ($U \simeq -37$~km/s, $V \simeq -17$~km/s). Those well-known overdensities
 (see Dehnen 1998, Chereul et al. 1999, Montes et al. 2001, Famaey et
 al. 2005) are either
 due to the evaporation of star clusters, or to non-axisymmetric
 perturbations by spiral arms and the galactic bar (e.g. the Hercules
 stream). Their prominence close to the center of the
 $UV$-plane is not compatible with the hypothesis that they are the result
 of merger events. Substantial amounts of such satellite debris have been
 identified in the Geneva-Copenhagen survey (Helmi et al. 2006), but at higher
 velocities and with lower metallicities.

\section{The Hyades stream}
We shall now focus our attention on the prominent Hyades
stream, which was suggested by Famaey et al. (2005) to be a dynamical stream coming
from the inner Galaxy because of a spiral perturbation, although firm proof was lacking because
metallicities were available for only very few stars and individual age
estimates were not precise. The main
question to be answered here is thus whether this kinematic group is really a
dynamical stream, or rather a coeval evaporated cluster.

An evaporated cluster would have a cluster-like homogeneity of age and
metallicity, while a dynamical stream would have the age and metallicity
composition of a random disk population (its metallicity could however
reflect that of the galactocentric radius wherefrom it originates as a
result of the dynamical perturbation).  

However, even with the high-accuracy data of the Geneva-Copenhagen survey, the 
precision on the metallicity and age is not high enough for one of the 
scenarios to stand out immediately. Indeed, the ages determined in
Nordstr\"om et al. (2004) from the position in the HR diagram cannot be
reliably used (Pont \& Eyer 2004): the uncertainties and systematic biases
on the age determination are strongly correlated with 
the mass, and a group of
stars with different masses and homogeneous age cannot be detected as such
from the computed ages (see Fig. 25 of Nordstr\"om et al. 2004).

Nevertheless, the two scenarios can be distinguished by a more subtle method
than the simple analysis of age and metallicity distributions.
If the Hyades stream is caused by a non-axisymmetric dynamical perturbation of the potential, we
expect a mass distribution similar to the field stars, obeying the present
day mass function (PDMF). On the other hand, if the Hyades stream is an evaporated cluster, 
composed of coeval stars (600 Myrs old), we expect a different mass
distribution, depending on the inital mass function
(IMF) of the cluster, and on the proportion of escaping stars as a
function of mass (see e.g. Terlevich 1987).

\subsection{Predictions in the purely coeval case}
\label{Sect:coeval}

The mass distribution thus offers us a good test of coevality for the
Hyades stream. Let us assume that the stream is coeval, entirely generated
by the on-going evaporation of the primordial Hyades cluster. We
consider two groups of stars, chosen to be as far apart as
  possible without suffering from selection biases and incompleteness 
(as illustrated by Fig.~\ref{Fig:byM}): 
\begin{itemize}
\item {\it Group 1}: stars with $1.3 \, M_\odot \leq m \leq 1.5 \,
  M_\odot$. The
  upper limit is below the Hyades cluster turnoff mass ($2.2 \, M_\odot$),
  so that no giant star from the assumed coeval stream is present in the
  group. To avoid contamination by the small number of field giants present
  in the Geneva-Copenhagen survey, we further restrict this group to $b-y <
  0.42$ (this supplementary condition does not alter the selection for
  main-sequence stars, see Fig.1). 

\item {\it Group 2}: stars with $0.8 \, M_\odot \leq m \leq 1 \,
  M_\odot$. 
\end{itemize}
Then, in this coeval scenario, the proportion of stars belonging to the
Hyades stream with respect to field stars of the disk, $N_{\rm hya}/N_{\rm
  field}$, should decrease from the value $p_1$ in
Group 1 to the value $p_2$ in Group 2, according to the variation of
the ratio of the mass function of the young evaporated Hyades cluster to
the PDMF of field stars. The estimation of the ratio $p_1/p_2$ under
  this coeval hypothesis nevertheless depends on the assumptions made on
  the efficiency of selective evaporation, namely on the proportion of
  evaporated stars from the cluster in the mass range corresponding to Groups 1 and 2. It is well-established that
  low-mass stars are preferentially depleted from star clusters from the
  effect of mass segregation (see e.g. Baumgardt \& Makino 2003), leading
  to a flattening of the observed cluster mass function, or even turning an
  initially increasing mass function into one which is decreasing towards
  low-mass stars (see e.g. Reid \& Hawley 1999, Dobbie et
  al. 2002). However, it is not at all obvious that, for a 600 Myrs open
  cluster, a significant preferential depletion should be expected for
  stars in the range $[0.8,1]$~$M_\odot$ as compared to stars in the range
  $[1.3,1.5]$~$M_\odot$.

\subsubsection{Without selective evaporation}

Let us first assume that the relative depletion of stars from the
primordial Hyades cluster is the same for the two mass groups under
consideration. The values $p_1$ and $p_2$ are then entirely determined by
the ratio of the Hyades cluster IMF to the field PDMF, integrated
over the appropriate mass intervals.

Moreover, since by construction Group 1 is restricted to dwarf stars, it
is also necessary to correct the PDMF for the absence of field giants when
estimating $p_1$: the ratio of the main sequence lifetime over total
lifetime is 80\% for a $1.5M_\odot$ star with $Z=0.008$ (Lejeune \& Schaerer
2001, assuming that the evolution posterior to the Helium flash represents
10\% of the MS lifetime), implying a correction of the PDMF by a factor 0.8
in Eq.(1) below. Such a correction is unnecessary for Group 2, dominated by
low-mass stars that merely have the opportunity to leave the main-sequence
when they are as old as the Galaxy itself.   

Adopting for the PDMF ${\rm d}N/{\rm d}m \propto m^{-4.5}$ (Kroupa et
al. 1993), and for the present-day
star formation IMF ${\rm d}N/{\rm d}m \propto m^{-2.3\pm0.7}$ for $1.3 \, M_\odot \leq m
\leq 1.5 \, M_\odot$, and
${\rm d}N/{\rm d}m \propto m^{-2.7\pm0.3}$ for $0.8 \, M_\odot \leq m \leq 1 \, M_\odot$ (Kroupa
2001), we have: 
\begin{equation}
\label{Eq:p1p2coeval}
\frac{p_1}{p_2} = \frac{\int_{1.3M_\odot}^{1.5M_\odot} {\rm IMF} \, {\rm
     d}m}{0.8 \int_{1.3M_\odot}^{1.5M_\odot} {\rm PDMF} \, {\rm d}m} \times \frac{\int_{0.8M_\odot}^{1M_\odot}
     {\rm PDMF} \, {\rm d}m}{\int_{0.8M_\odot}^{1M_\odot}
     {\rm IMF} \, {\rm d}m} \, = 3.18_{-0.72}^{+1.03}
\end{equation}

\subsubsection{With selective evaporation}

N-body simulations of the dynamical evolution of open clusters with an initial
number of 1000 stars (Terlevich 1987) have long ago demonstrated that
evaporation was more effective for low-mass stars because of mass
segregation. This trend is extremely clear in numerical simulations if one
compares the relative depletion of stars lighter and heavier than $m \sim 0.5$ $M_\odot$. However,
if one compares the relative number of escapers from our two mass groups
after 75\% of the stars have left the cluster, the effect is not clear at
all (see Fig. 2 of Terlevich 1987). The numerical noise is high, and the
depletion is favoured in one group or the other depending on the initial
conditions. The simulations seem to indicate that
the relative number of escapers in Group 2 could be higher than in Group 1
{\it at the very most} by a factor $\kappa = 1.4$. While such an efficient
selective evaporation would {\it steepen} the observed mass function of stars
evaporated from the
cluster to bring it closer to the PDMF (and would {\it decrease} the
predicted ratio $p_1/p_2$ by a factor $\kappa$), it would actually also
{\it flatten} the observed mass function  of the cluster itself. From
Fig. 8 of Reid \& Hawley (1999), it appears that (for the mass range under
consideration here), the Hyades cluster MF has a power-law exponent $\sim
-2.3$. A  factor $\kappa$ for the relative number of escapers in the range
$-0.1<{\rm log}(m)<0$ as compared to the relative number of escapers in the
range $0.1<{\rm log}(m)<0.2$ would then lead to an IMF with
a power-law exponent of $-2.3 - 5 \, {\rm log}(\kappa)$. An extreme
scenario, combining a selective evaporation with a maximal $\kappa=1.4$ and
the steepest IMF $\propto m^{-3.0}$ (allowed by Kroupa 2001 error bars) is
thus not formally excluded. In that extreme case, the lower bound in Eq.(1) 
(corresponding to an IMF power-law exponent of $-3.0$) should be
divided by $\kappa=1.4$, thus yielding $p_1/p_2 = 1.76$. 



\subsection{Observational test of coevality}

  \begin{figure}
   \centering
   \includegraphics[width=8cm]{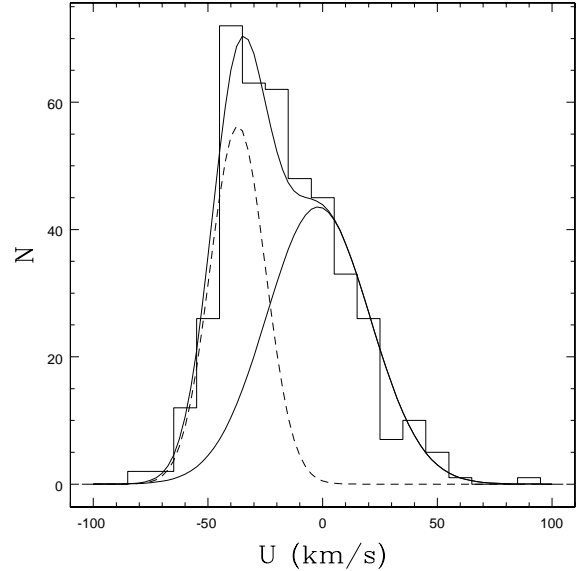}
      \caption{Histogram of the $U$ velocities for the 415 stars in Group 1,
with $-21 \le V ({\rm km/s}) \le -12$. The dashed line corresponds to
the contribution of the Hyades stream.}
         \label{histoU_group1}
   \end{figure}

  \begin{figure}
   \centering
   \includegraphics[width=8cm]{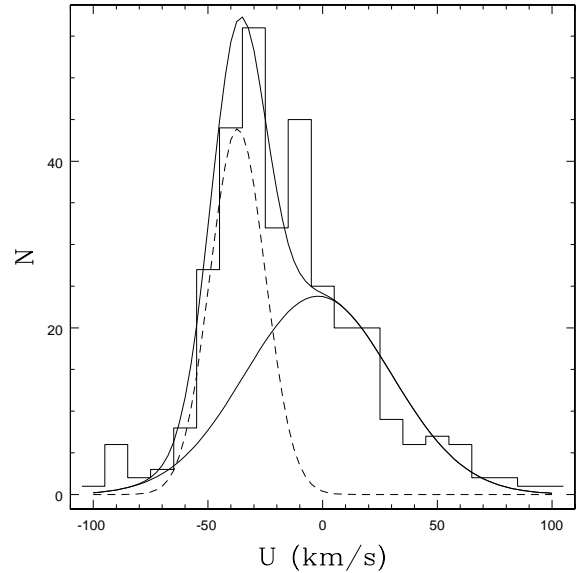}
      \caption{Same as Fig.~\protect\ref{histoU_group1} for the 323 stars
      in Group 2. The sharp excess in the $U$-bin $[-15,-5]$ is due to
      the presence of the Castor moving group, and to a high relative Poisson
      noise.} 
         \label{histoU_group2}
   \end{figure}

Let us now determine the observed value of $p_1$ and $p_2$ from the
Geneva-Copenhagen survey. In order to estimate $N_{\rm hya}/N_{\rm field}$
in the two mass groups, we need to separate the Hyades stream from the
field of the galactic disk. We therefore consider the strip $-21 \leq V
({\rm  km/s}) \leq -12$  in the $UV$-plane, corresponding to the range in
$V$ covered by the Hyades stream (see Fig.~2). We only consider stars with
[Fe/H]~$>-0.5$ to avoid contamination by the halo (the Hyades
overdensity is {\it not seen} in the $UV$-plane for stars with
[Fe/H]~$<-0.5$). 
 
Figs.~3 and 4 display the histogram along $U$
in the $V$-strip for Groups~1 and 2, respectively. The relative Poisson
noise is larger for Group~2 due to the smaller number of
stars. 
Neglecting the Castor moving group appearing as a sharp peak 
in the $U$-bin $[-15,-5]$
in Group 2, these histograms may be represented by a superposition of
two normalized gaussians ${\cal N}(\langle U \rangle, \sigma_U)$, the
first one corresponding to the Hyades and the second
one to the field of the galactic disk: 
\begin{equation}
\label{Eq:k1}
\tilde{N}_1(k_1) = \overline{N}_1\; [\;(1-k_1) \times {\cal N}(-37,12) + k_1 \times {\cal N}(-2,22.5)\;]
\end{equation}
for Group 1, and
\begin{equation}
\label{Eq:k2}
\tilde{N}_2(k_2) = \overline{N}_2\; [\;(1-k_2) \times {\cal N}(-37,12) + k_2 \times {\cal N}(-2,32)\;]
\end{equation}
for Group 2, where $\overline{N}_1, \overline{N}_2$ are the total number of stars of
the considered group in the $V$-strip, multiplied by the width of the
$U$-bins, and $k_1, k_2$ the fraction 
belonging to the field. The mean value $\langle U \rangle =
-2$ km/s for field stars does correspond to the value found for
giant stars
in Famaey et al. (2005) when the streams were not taken into account. The
velocity dispersion of the field increases with decreasing mass ($\sigma_U
= 22.5$ km/s in Group~1 and $\sigma_U = 32$~km/s in Group~2), as expected
from the age-velocity dispersion relation (see e.g. Dehnen \& Binney 1998).

The parameters $k_1$ and $k_2$ have been derived numerically 
(separately for Group~1 and 2: $j = 1,2$) by minimizing:
\begin{equation}
\label{Eq:chi2}
\chi^2(k_j) = \Sigma_{i=1}^M \frac{[\;N_i - \tilde{N}_i(k_j)\;]^2}
{\tilde{N}_i(k_j)},
\end{equation}
where the sum extends over the $M$ bins in $U$, $N_i$ is the
observed number of stars in bin $i$, and the denominator $\tilde{N}_i$ 
corresponds to the expected number. 
In Group~2, the bin containing Castor has not been included in the sum. Pearson's $\chi^2$ 
goodness-of-fit criterion used here should only be applied to bins containing at least 5 stars. 
Therefore, 
the fitting process has been restricted to the range $-65 \le
U(\mathrm{km/s}) \le 55$.  

The uncertainties $\sigma_{k_1}, \sigma_{k_2}$ 
on the parameters $k_1, k_2$ (listed in Table~\ref{Tab:Vstrip}) 
have been estimated from
the relation
\begin{equation}
\sigma^2_{k_j} = \Sigma_{i=1}^M\; N_i\; \left( \frac{\partial k_j}{\partial
  N_i} \right)^2, 
\end{equation}
where the derivatives $\partial k_j / \partial N_i$ are derived from the
condition $\partial \chi^2 / \partial k_j = 0$, yielding
\begin{equation}
\sigma^2_{k_j} =  \left(\Sigma_{i=1}^M\; \frac{N_i\;}{\left[\;k_j + \frac
    {{\cal N}(-37,12)}{{\cal N}(-2,22.5) \;-\;  {\cal N}(-37,12)}\;\right]^2} \right)^{-1}
\end{equation}
under the assumption that the number of stars in the bins are large
enough to ensure that $|N_i - \tilde{N}_i|/\tilde{N}_i << 1$.

In order to derive $p_1$ and $p_2$, we need to estimate the proportion of field
stars in the considered strip $-21 \leq V({\rm km/s}) \leq -12$ with respect
to the whole $UV$-plane. To avoid contamination by other streams present in our
sample, we simply approximate  the
field-star $V$-distribution by ${\cal
  N}(-10,12)$ for Group~1 and ${\cal N}(-15,18)$ for Group~2, according to
the age-velocity dispersion relation and the asymmetric drift (Dehnen
\& Binney 1998). We thus obtain  
\begin{equation}
\label{Eq:p1}
p_1 = (1-k_1)/k_1 \times \int_{-21}^{-12} {\cal N}(-10,12)\, {\rm
     d}V = 0.175\pm0.024,
\end{equation}
and
\begin{equation}
\label{Eq:p2}
p_2 = (1-k_2)/k_2 \times \int_{-21}^{-12} {\cal N}(-15,18)\, {\rm
     d}V = 0.136\pm0.025,
\end{equation}
where the errors have been propagated according to $\sigma_p =
  \sigma_k \;  |{\rm d}p/{\rm d}k| = \sigma_k  / k^2
  \int_{-21}^{-12} {\cal N} \, {\rm d}V$. 
We thus have:
\begin{equation}
\label{Eq:value}
\frac{p_1}{p_2} = 1.29_{-0.35}^{+0.50}.
\end{equation}
In order to evaluate the sensitivity of these results to the choice of
the $V$ strip ([$-21, -12]$ so far), Table~\ref{Tab:Vstrip} lists
$k_1, k_2, p_1, p_2$ for slightly different choices of this strip. All
these choices actually yield very similar results for the $p_1/p_2$ ratio, clearly incompatible with 
the standard coeval scenario without
selective evaporation, predicting $3.18_{-0.72}^{+1.03}$ [see Eq.(1)].
\begin{table*}
\caption[]{\label{Tab:Vstrip}
Sensitivity of the parameters characterizing the relative fraction of
Hyades and field stars to the choice of the $V$ strip enclosing the Hyades
(see Eqs.~\ref{Eq:k1}, \ref{Eq:k2}, \ref{Eq:chi2}, \ref{Eq:p1} and \ref{Eq:p2} 
for the definition of the various parameters listed). For Group~1, 
Pearson's goodness-of-fit $\chi^2(k_1)$ behaves roughly like 
a $\chi^2$ distribution with 10 degrees of freedom (12 bins - 2 constraints). For Group~2, that number
reduces to 9 (because of the exclusion of Castor's bin).  
The probability that a random variable exceeds the observed $\chi^2$ value is in all cases large 
enough for the two-gaussion model to be considered as a satisfactory fit to the data 
[To fix the ideas, Prob$(\chi^2_{10} > 17.4) = 6.6$\% and Prob$(\chi^2_{9} > 13.7) = 13.4$\%]. 
}
\begin{tabular}{cccccccc}
\hline
\hline
$V$ strip & $k_1$ & $\chi^2(k_1)$ & $k_2$ & $\chi^2(k_2)$ & $p_1$ & $p_2$ & $p_1/p_2$ \\
(km/s)\\
\hline
{[}-21, -12] & $0.592\pm0.033$ & 13.8 & $0.591\pm0.045$ & 11.3 & $0.175\pm0.024$ &
$0.136\pm0.025$ & $1.29_{-0.35}^{+0.50}$ \\
{[}-20, -13] & $0.572\pm0.037$ & 17.4 & $0.566\pm0.051$ & 11.0 & $0.190\pm0.029$ &
$0.151\pm0.031$ & $1.26_{-0.37}^{+0.57}$ \\
{[}-22, -13] & $0.588\pm0.033$ & 16.6 & $0.606\pm0.046$ & 9.9  & $0.178\pm0.024$ &
$0.128\pm0.025$ & $1.39_{-0.38}^{+0.57}$ \\
{[}-20, -11] & $0.595\pm0.034$ & 14.3 & $0.576\pm0.046$ & 13.7 & $0.173\pm0.024$ &
$0.145\pm0.027$ & $1.19_{-0.33}^{+0.48}$ \\
\hline
\end{tabular} 
\end{table*}
Even the most favourable coeval scenario, with a steep IMF of power-law
  exponent $-3.0$, and with a significant preferential evaporation of 1
  $M_\odot$ stars over 1.5 $M_\odot$ stars,
  yields $p_1/p_2 = 1.76$, a ratio higher than the observed
  value by one sigma. In this scenario, the
  mass of the stream (around $800$~$M_\odot$ in the Geneva-Copenhagen
  catalogue alone, i.e. a lower bound since it contains only FG dwarfs)
  implies a very massive primordial Hyades cluster. Chumak et
  al. (2005) showed that, after 600 Myrs, $800$~$M_\odot$ could be
  lost by
  the cluster if the initial mass was $1400$~$M_\odot$ and the initial virial
  radius 7.5 pc: this evaporated mass would be distributed in a volume of
  radius 300 pc around the Sun, but since $800$~$M_\odot$ is observationally
  a lower bound, even more extreme initial conditions would probably be
  needed to account for the full Hyades stream. We thus conclude that 
  the most likely scenario is that the Hyades overdensity in the $UV$-plane 
  is {\it not uniquely} composed of coeval stars evaporated from the primordial
  Hyades cluster.    

Note that the zeroth order expectation $p_1=p_2$ for a stream exclusively
composed of field stars is well within the error bars of
Eq.(\ref{Eq:value}). However, the best fit value yields $p_1>p_2$. There
might be several reasons for this:
\begin{itemize}
\item First of all, since the ratio of Eq.(\ref{Eq:value}) was
  observationally estimated under the coevality hypothesis, we did not take
  into account possible variations of the intrinsic velocity dispersion of
  the Hyades stream itself. However, under the dynamical perturbation
  hypothesis, the Hyades stream itself could be affected by the
  age-velocity dispersion relation. For instance, if we take for the Hyades
  $\sigma_U = 15$~km/s instead of $12$~km/s in the older Group 2
  (Eq.\ref{Eq:k2}), we find $p_1/p_2=1.15_{-0.32}^{+0.47}$, a value closer
  to one than Eq.(\ref{Eq:value}).

\item $p_1>p_2$ might be caused by the PDMF of the Hyades stream being slightly
  different from the PDMF of the solar neighbourhood. As we shall see in
  the next section, the Hyades stream is over metal-rich at all masses,
  pointing towards an origin in the inner Galaxy. Since the lifetime of
  stars depends on the metallicity, the PDMF of the inner Galaxy should be
  slightly different from the PDMF of the solar neighbourhood, even with
  identical IMF and stellar formation rates (SFR). 
Moreover, the SFR could also vary as a function of the
  galactocentric radius.

\item $p_1>p_2$  might be due to {\it some} of the stars in the stream still
  belonging to the evaporated cluster since we know that the
Hyades cluster is kinematically associated with the Hyades stream, and we
also know that this cluster does evaporate with time and must create a velocity
clump of spatially extended stars. Assuming the same PDMF
for the Hyades stream and the solar neighbourhood (zeroth order
  approximation), and the best fit value for the ratio $p_1/p_2$, we can
  actually estimate what proportion of stars in each group is associated with the coeval Hyades cluster, and what proportion
  is associated with the dynamical stream. Eq.~(1) gives the ratio of the
  evaporated cluster proportion in Group 1 with respect to this proportion
  in Group 2, while the ratio is assumed to be 1 for the dynamical
  stream. Eqs.~(\ref{Eq:p1}) and (\ref{Eq:p2}) then give the sum of these
  proportions in each group. From there we find that the evaporated cluster
  represents about 40\% of the total stream in Group 1, but only 15\% in
  Group 2.  These numbers are given here as an indication since these
  proportions should be affected by the variation of the stream PDMF, and
  could of course be zero within the error bars of Eq.(\ref{Eq:value}).
\end{itemize}

\subsection{Metallicity of the stream}
We have thus shown that the vast majority of stars in the
Hyades stream are most likely field-like disk stars obeying the PDMF of the
disk. This does not necessarily imply that the stream originates from a
non-axisymmetric perturbation of the Galactic potential. One could instead
imagine that the primordial Hyades open cluster was an extremely massive
object able to trap some older galactic field stars during its
formation process (see e.g. Fellhauer et al. 2006). These stars could have
formed a dynamically hotter sub-system in the outer part
of the object, and could have been preferentially evaporated (whilst
obeying the PDMF). However, in that case, one would expect the stream to
exhibit a slightly sub-solar metallicity, characteristic of the solar
galactocentric radius, close to which the Hyades cluster is supposed to
have formed if it has not been dynamically perturbed by spiral arms. 

The mean metallicity of the Geneva-Copenhagen survey is $[Fe/H] =
-0.16$. Excluding halo stars with $[Fe/H]<-0.5$, and excluding all the stars
from the Hyades box ($-21$ km/s $\leq V \leq -12$ km/s and $-50$ km/s $\leq
U \leq -25$ km/s), we get a mean disk metallicity of $[Fe/H]=-0.126 \pm
0.002$ for the solar neighbourhood (the error representing the standard
error of the mean).

If we rather concentrate on the Hyades box, we get $[Fe/H]=-0.061 \pm
0.012$ for Group 1, and $[Fe/H]=-0.059 \pm 0.013$ for Group 2, meaning that
{\it the stream is over metal-rich at all masses}. We can actually
calculate the ratio of stars from the stream to stars from the field disk
in the Hyades box from Figs. 3 and 4. We get $N_{\rm hya}/N_{\rm
  field}({\rm box})=3.65$ for Group 1,
and $2.88$ for Group 2, meaning that the stream has a roughly constant mean
metallicity $[Fe/H]=+0.02$, even slightly rising for low-mass stars.
This is inconsistent with the hypothesis of a field population from the solar
neighbourhood trapped in the primordial cluster and subsequently evaporated.  

On the other hand, this constancy of the metallicity is consistent
  with the assumption that evaporated and field-like stars in the
  stream have roughly the same metallicity. This could imply that they {\it
  both} come from the inner regions of the Galaxy after having crossed the
  corotation of the spiral pattern (the cluster could have been shifted in
  radius while remaining bound since the effect of a spiral wave on stars
depends on the stars' phase with respect to the spiral, and the phase
does not vary much across the cluster). The metallicity excess of 0.15 dex
  is indeed not consistent with the present-day
orbit of the Hyades stream: this present-day orbit is centered on a
guiding radius $R_g \sim 7.5$ kpc in galactocentric coordinates ($8$ kpc being
the galactocentric radius of the Sun), and would imply an implausible
metallicity gradient of -0.3 dex/kpc in the disk. Assuming a very steep but
  realistic (see e.g. Daflon \& Cunha 2004) galactic metallicity gradient of -0.07 dex/kpc,
the metallicity excess is compatible with a galactocentric origin near
$R=6$ kpc. This would be consistent with the order of magnitude of stellar
  wandering caused by spiral perturbations (2-3 kpc, see Sellwood \& Binney
  2002; L\'epine et al. 2003). A flatter metallicity gradient would imply
  an even more internal origin for the stream. 

\section{Discussion}

Using the Geneva-Copenhagen catalogue of F and G dwarfs (Nordstr\"om et
al. 2004), we have analyzed the mass function of the Hyades stream, an
overdensity of stars in the $UV$-plane kinematically associated with the
Hyades cluster.  

In order to be compatible at a one-sigma level with the predicted mass
function for stars evaporated from the primordial Hyades cluster, one needs
a rather extreme scenario, with a very significant preferential evaporation
from the cluster of 1~$M_\odot$ stars over 1.5~$M_\odot$ stars, and with a
very steep IMF, not flatter than a power-law exponent of -3.0 (see
Sects. 4.1 and 4.2). The initial total mass and virial radius of the
cluster should also be very high ($M>1400$ $M_\odot$, $r>7.5$ pc).  

On the other hand, the observed mass function (see Sect. 4.2,
Eq.~\ref{Eq:value}) is perfectly compatible with the hypothesis that the
stream is uniquely composed of field-like stars obeying the PDMF. However,
because the Hyades cluster is known to share the kinematics of the stream
and to evaporate over time, the most likely scenario is that the stream is
indeed mainly composed of
field-like stars (about 85\% of the stream for low-mass stars), but also partly
of coeval stars evaporated from the primordial Hyades cluster (about 15\%
of the stream for low-mass stars). 

The peculiar metallicity of the
stream at all masses (see Sect. 4.3) is not
compatible with a scenario where the field-like stars would have been
trapped in the primordial Hyades cluster during its formation process, and
subsequently evaporated. Those stars were thus most probably trapped at
resonance by a spiral perturbation. Indeed, a series of strong transient
spirals with their mean corotation at the solar galactocentric radius are
known to produce small-scale structure in the local velocity distribution
(De Simone et al. 2004). Stars on horseshoe orbits that cross the
corotation could wander over 2-3 kpc in much less than 1 Gyr (Sellwood \&
Binney 2002; L\'epine et al. 2003), while staying on quasi-circular orbits,
not elongated enough to betray their place of birth. This is consistent
with a galactocentric origin near $R=6$ kpc for the Hyades stream, that
could explain its metallicity excess at low masses. On the other hand, the
Hyades stream could also correspond to nearly closed orbits trapped at the
$4:1$ inner Lindblad resonance of a two-armed spiral density wave (Quillen
\& Minchev 2005). In any case, the prominence of the Hyades dynamical
stream raises the question of the amplitude of the spiral perturbation
needed to produce such a stream. The most detailed models of gas flows in
the Galaxy (Bissantz et al. 2003) indicate that the amplitude of the
spiral structure in the mass density is larger by a factor $1.5$ than its
amplitude in the near-infrared luminosity density, while such a large
amplitude is also needed to produce the large non-axisymmetric motion of the
star-forming region W3OH (Xu et al. 2006). This requires the
baryonic disk to be massive even near the Sun, and is very constraining for
the dark matter distribution in the Galaxy (see e.g. Famaey \& Binney
2005). On the other hand, the probable resonant origin of the Hyades
stream cautions against naive backward integration of individual orbits in
an axisymmetric potential when describing the past evolution of such
streams, and it bares the importance of evaluating the impact of radial
migrations on the chemical evolution of the Galaxy (see also Haywood 2006). 

\begin{acknowledgements}
We thank the referee Michel Cr\'ez\'e for many helpful suggestions that
have considerably improved the present paper. BF is FNRS Research
Associate, AJ is FNRS Senior Research Associate. XL acknowledges the
support from the Spanish MCyT through grant AYA2003-07736.  
\end{acknowledgements}

\end{document}